\documentclass[12pt,reqno]{amsart}
\usepackage[latin1]{inputenc}

\newcommand{\G}{\Gamma}

\newcommand{\C}{\mathbb C}

\newtheorem{theorem}{Theorem}[section]
\newtheorem{lemma}[theorem]{Lemma}
\newtheorem{proposition}[theorem]{Proposition}

\theoremstyle{definition}
\newtheorem{definition}[theorem]{Definition}

\theoremstyle{definition}

\newtheorem{remark}[theorem]{Remark}

\begin{document}

\title[\hfilneg {        }\hfil      higher-order $q$-Bessel
heat equation    ] {Polynomial Expansions for Solutions of \\
higher-order $q$-Bessel heat equation}

\author{M.S.BEN HAMMOUDA }

\address{Med Saber Ben Hammouda \newline
D\'epartement de Mat\'ematiques, Facult\'e Des Sciences De Tunis,
Campus Universitaire D\'el Menzah, Tunis 1060, Tunisie }
\email{Saber.BHMED@fst.rnu.tn}

\author{Akram Nemri}

\address{Akram Nemri \newline
D\'epartement de Mat\'ematiques, Facult\'e Des Sciences De Tunis,
Campus Universitaire D\'el Menzah, Tunis 1060, Tunisie }
\email{Akram.Nemri@fst.rnu.tn}

\keywords{ $q$-Analysis, $q$-Fourier transform, $q$-Heat equation,
$q$-Laguerre polynomials, $q$-Heat polynomials . \hfill\break}

\begin{abstract}
In this paper we give the $q$-analogue of the higher-order Bessel
 operators studied by  M. I. Klyuchantsev \cite{M.I.Klyu}
and A. Fitouhi, N. H. Mahmoud and S. A. Ould Ahmed Mahmoud
\cite{FIT2}. Our objective is twofold. First, using the
$q$-Jackson integral and the $q$-derivative, we aim at
establishing some properties of this function with proofs similar
to the classical case. Second our goal is to construct the
associated $q$-Fourier transform and the $q$-analogue of the
theory of the heat polynomials introduced by P. C. Rosenbloom and
D. V. Widder \cite{RW1}. Our  operator for some value of the
vector index
 generalize the $q$-$j_{\alpha}$ Bessel operator of the second order
in \cite{FIT3} and a $q$-Third operator in \cite{FIT sab}.

\end{abstract}
\maketitle

\section{Introduction}

The Bessel operator of $r$-order is defined on $(0, \infty)$ by

\begin{equation}
   B_r u = u^{(r)} +
\frac{a_1 }{x}u^{(r - 1)} + ... + \frac{a_{r - 1} }{x^{r -
1}}u^{(1)},
\end{equation}
where the coefficients $a_{k}$ depend on the components $\alpha _k$
\begin{equation}\alpha _k \ge - 1 + \frac{k}{r} \quad , \qquad k =
1,...,r - 1.\end{equation} and
\begin{equation}
a_{r - k} = \frac{1}{(k - 1)!}\sum\limits_{j = 1}^k {( - 1)^{k - j}
\binom{j - 1}{k - 1} } \prod\limits_{i = 1}^{r - 1} {(r\alpha _i +
j)}.
\end{equation}

Where $r$ is positive integer and  $\alpha = (\alpha _1 ,...,\alpha
_{r - 1} )$ a vector having $(r-1)$ components with $\left| \alpha
\right| = \alpha _1 + ... + \alpha _{r - 1}$.

When $r = 2$, we obtain the classical Bessel operator of the second
order
\begin{equation}
B_2 u = u'' + \frac{2\alpha + 1}{x}u'\, ,
\end{equation}

and for $r = 3$, $\alpha _1 = -2/3$, $\alpha _2=\nu-1/3$, we obtain
the operator $B_{3} u$ studied in \cite{CholH68} and in \cite{FIT
sab1}
\begin{equation}
B_{3} u =\frac{d^3}{dx^3} + \frac{3\,\nu}{x}\,\frac{d^2}{dx^2} -
\frac{3\,\nu}{x^2}\,\frac{d}{dx}, \qquad \nu> 0.
 \label{equ12006}
\end{equation}

For $\lambda$ being a complex number, let us now consider the system
\begin{align}
\begin{cases}
B_{r}u(x) & =\quad  -\lambda^{r} u(x),\\ \quad u(0)&=\quad 1,\;
\\ u^{k}(0)&=\quad  0,\quad k=1,...,r-1.
\end{cases}\label{equ22006}
\end{align}
The use of the Frobenius method leads us to conclude that
(\ref{equ22006}) has a unique solution which is $r$-even and given
by

\begin{equation}\label{equ32006}
j_\alpha (\lambda x) = \sum\limits_{m \ge 0} {( - 1)^m
\frac{1}{m!}\prod\limits_{i = 1}^{r - 1} {\frac{\Gamma (\alpha _i +
1)}{\Gamma (\alpha _i + m + 1)}} \left( {\frac{\lambda x}{r}}
\right)^{rm}  }.
\end{equation}

In this paper we are concerned with the $q$-analogue of the
$j_\alpha $ higher-order Bessel function (\ref{equ32006}). This
choice is motivated in particular by the context of \{\cite{FIT2},
\cite{FIT3}, \cite{FIT sab}\}.

The reader will notice that the definition (\ref{j}) derives from
that given in \cite{FIT2} with minor changes. With the help of the
$q$-integral representation we establish the $q$-integral
representation of the Mehler and Sonine types. Moreover, we define
the higher-order $q$-Bessel translation and the higher-order
$q$-Bessel Fourier transform and  establish easily some of their
properties. Finally, we study the higher-order $q$-Bessel heat
equation.


\section{ Notation and Preliminary Results}

Let $q$ be a fixed real number $0<q<1$. Henceforth, we use the
following notation:
\begin{eqnarray}
(a+b)_q^n = \prod_{j=0}^{n-1}(a+q^jb), \text{ if } n =0, 1, 2, ...
,\infty, \quad \, (1+a)_q^t =
\frac{(1+a)_q^\infty}{(1+q^ta)_q^\infty},
  \text{ if } t\in\C\ . \label{intr4}
\end{eqnarray}
We note for $\lambda \in \mathbb{R}_{q},\, n =0, 1, 2, ... ,$
\begin{align}
(a;\,q)_{n} =& (1-a)(1-aq)...(1-aq^{n-1}),\\
(\lambda)_{q}=\frac{1-q^{\lambda}}{1-q},& \qquad
(\lambda)_{n}^{q} =\frac{(q^{\lambda};\,q)}{(1-q)^{n}},\quad [n]_{q}! = \frac{(q;\,q)_{n}}{(1-q)^{n}}, \\
  \frac{(\lambda)_{n}^{q}}{[n]_{q}!} =   \frac{(q^{\lambda};\,q)_{n}}{(q;\,q)_{n}}, &   \quad  \frac{(\lambda)_{n}^{q}}{(\lambda + n - 1)_{q}} = (\lambda)_{n-1}^{q} ,   \quad  \frac{(1)_{n}^{q}}{(1)_{n-k}^{q}}=(-1)^{k}(-n)_{k}^{q}q^{nk-\binom{k}{2}}.
\end{align}

\subsection{The $q$-Binomial formula } We note a $q$-Binômial formula by  :
\begin{equation}\label{binom}
(ab;\,q)_{n} = \sum_{k=0}^{n}\left[\begin{matrix} {n} \\
{k}\end{matrix}\right]_{q} b^{k}(a;\,q)_{k} (b;\,q)_{n-k}, \\
\end{equation}
with
\begin{equation}
 \left[\begin{matrix}{n} \\
{k}\end{matrix}\right]_{q} =
\frac{(q;\,q)_{n}}{(q;\,q)_{n-k}(q;\,q)_{k}  },\quad n\in
\mathbb{N},\, k = 0,1,...,n.
\end{equation}

\subsection{The $q$-derivative and the $q$-integral }
We denote by $D_q$ the $q$--derivative of a function
\begin{equation}\label{intr9q}
D_q f(x)=\frac{f(qx)-f(x)}{(q-1)x}\ .
\end{equation}

\begin{equation}\label{intr10pp}
  D_{q}^{n}f(x) = \frac{q^{-\binom{n}{2}}}{x^{n}
  (1-q)^{n}}\sum_{k=0}^{n}(-1)^{k}\left[\begin{matrix}{n} \\
{k}\end{matrix}\right]_{q} q^{\binom{n-k}{2}}f(q^{k} x ),\;n =
0,1,2,....
\end{equation}
\begin{equation}\label{intro11}
   D_{q}^{n}[f(x) g(x)] = \sum_{k=0}^{n}\left[\begin{matrix}{n} \\
{k}\end{matrix}\right]_{q}(D_{q}^{n-k}f)(q^{k}x)(D_{q}^{k}g)(x), \;n
= 0,1,2,....
\end{equation}

We define the $q$-schift operators by :
\begin{align}
(\Lambda_{q}f)(x) = f(qx)  & \qquad and  \qquad
(\Lambda^{-1}_{q}f)(x) = f(q^{-1}x),\\ D_{q}\Lambda_{q} = q
\Lambda_{q}D_{q} & \qquad and \qquad D_{q}\Lambda_{q}^{-1}= q^{-1}
\Lambda_{q}^{-1} D_{q}.
\end{align}
and also we note  $(\Lambda^{-1}_{q^{\delta}}f)(x) =
f(q^{-\delta}x)$.

The $q$-Jackson integrals (introduced by Thomae  and Jackson
\cite{jackson_1910}) from $0$ to $a$ and from $a q$ to $\infty$ are
defined by
\begin{equation}\label{intr8}
\int_0^{a}f(x)d_qx =(1-q)\sum_{j=0}^\infty aq^jf(aq^j) \text{ and }
  \int_{aq}^{\infty}f(t) d_{q}t = (1-q)\sum_{k=0}^{+\infty} a
  q^{-k}f(aq^{-k}).
\end{equation}
Notice that the series on the right hand side are guaranteed to be
convergent. See \cite{FIT3}.
\\

We define the Jackson integral in a generic interval $[a,b]$ by
\cite{jackson_1910}: $$
\int_a^bf(x)d_qx=\int_0^bf(x)d_qx-\int_0^af(x)d_qx\ . $$

This is a special case of the following more general change of
variable formula, \cite[p 107]{kac_qc}. If $u(x)=\alpha x^\beta$,
then $ \int_{u(a)}^{u(b)}f(u)d_qu
=\int_a^bf(u(x))D_{q^{1/\beta}}u(x)d_{q^{1/\beta}}x\ . $

Using the $q$-Jackson integrals from $0$ to $1$, we define the
$q$-integral $\int_{0}^{1}...\int_{0}^{1}
f(t_{1},...,t_{n})d_{q}t_{1}...d_{q}t_{n}$ by
\begin{equation}\label{intr80} \int_{0}^{1}...\int_{0}^{1} f(t_{1},...,t_{n})  \,
d_{q}t_{1}...d_{q}t_{n} =(1-q)^{n} \,
\sum_{i_{1},...,i_{n}=0}^\infty
q^{i_{1}+...+i_{n}}\;f(q^{i_{1}+...+i_{n}}),
\end{equation}
provided the sums converge absolutely.

\subsection{$q$-Exponential function}
We present two $q$--analogues exponential function:

\begin{eqnarray}
E_q(x) &=& \sum_{n=0}^\infty q^{\binom{n}{2}}\frac{x^n}{[n]_{q}!}\
=\ (1+(1-q)x)_q^\infty\ , \label{intr6} \\ e_q(x) &=&
\sum_{n=0}^\infty \frac{x^n}{[n]_{q}!}\ =\
\frac{1}{(1-(1-q)x)_q^\infty}\ . \label{intr7}
\end{eqnarray}
 Notice that for $q\in(0,1)$ the series
expansion of $e_q(x)$ has radius of convergence $1/(1-q)$. On the
contrary, the series expansion of $E_q(x)$ converges for every $x$.
Both product expansions (\ref{intr6}) and (\ref{intr7}) converge for
all $x$.

\subsection{$q^{\delta}$-Basic hypergeometric series }
We define the $q^{\delta}$-basic hypergeometric series
${_{r}\phi_{s}^{\delta}}$ by
\begin{equation*}
{_{r}\phi_{s}^{\delta}}\left(\begin{matrix} q^{a_{1}},...,q^{a_{r}}
\\ q^{b_{1}},...,q^{b_{s}}
\end{matrix}\bigg|q;(q-1)^{1+s-r}z \right)=
\sum_{k=0}^{\infty}(q^{\delta})^{\binom{k}{2}}\frac{(a_{1};q)_{k}^{q}...(a_{r};q)_{k}^{q}
}{(b_{1};q)_{k}^{q}...(b_{s};q)_{k}^{q} }\frac{z^{k}}{[k]_{q}!} ,
\end{equation*}
\begin{equation}
\lim_{q\uparrow1}{_{r}\phi_{s}^{\delta}} \left(\begin{matrix}
q^{a_{1}},...,q^{a_{r}} \\ q^{b_{1}},...,q^{b_{s}}
\end{matrix}\bigg|q;(q-1)^{1+s-r}z \right)= _{r}F_{s}\left[
\begin{matrix}
a_{1},...a_{r} \\ b_{1},...b_{s}
\end{matrix} \bigg| z \right].
\end{equation}
Where ${\delta}
> 0$ and $r < s + 1$, this expansion converges for all values of $z$ .

For ${\delta}=  1 + s - r$, we obtain the classic basic
hypergeometric series ${_{r}\phi_{s}}$\ , \cite[p 11,12]{koekoek}.

We note for ${\delta} > 0$ by $e_q(x , \delta) =\displaystyle
\sum_{n=0}^{\infty}\,q^{\delta
\binom{n}{2}}\,\frac{x^{n}}{[n]_{q}!}$, this expansion converges for
all values of $x$ .
\subsection{$q$--Gamma and $q$--Bêta functions}
The $q$--gamma function $\G_q(t)$, a $q$--analogue of Euler's gamma
function, was introduced by Thomae  and later by Jackson  as the
infinite product
\begin{equation}\label{int1}
\G_q(t)=\frac{(1-q)_q^{t-1}}{(1-q)^{t-1}}\ ,\quad t>0\ .
\end{equation}

The $q$--Beta function defined by the usual formula
\begin{equation}\label{intr9}
\beta_q(t,s)=\frac{\G_q(s)\G_q(t)}{\G_q(s+t)}\ ,
\end{equation}
has the $q$--integral representation, which is a $q$--analogue of
Euler's formula:
\begin{equation}\label{intr10}
\beta_q(t,s)=\int_0^1x^{t-1}(1-qx)_q^{s-1}d_qx\ ,\quad t,s>0\ .
\end{equation}


The $q$-duplication formula holds
\begin{equation}\label{001}
 \prod_{i=1}^{r-1}
\Gamma_{q^{r}}(n + \frac{i}{r}) \, \frac{1}{[r n]_{q}!} =
\prod_{i=1}^{r-1}  \Gamma_{q^{r}}( \frac{i}{r}) \,
\frac{1}{[n]_{q^{r}}!}\,\frac{1}{{((r)_{q}})^{rn}},
\end{equation}

and \begin{equation}
{((r)_{q})^{rn}(1)_{n}^{q^{r}}{\prod_{i=1}^{r-1}}(\frac{i}{r})_{n}^{q^{r}}}
=  {[r n]_{q}!}.
\end{equation}

We also denote, \quad $\displaystyle {\prod }(\alpha_{i} +1
)_{n}^{q^{r}} = {\prod_{i=1}^{r-1}}(\alpha_{i} +1 )_{n}^{q^{r}}. $

\section{$q$-Trigonometric Fonction of $r$-order}

The $r-q^{\delta}$-cosinus is defined for $\delta > 0$ by

\begin{align}\label{0001}
\cos_{r}(x,\, q^{r};{\delta})&
={_{0}\phi_{r-1}^{\delta}}\left(\begin{matrix} -
\\ (q^{r})^{1/r},...,(q^{r})^{(r-1)/r}
\end{matrix}\bigg|q^{r};- \frac{(q^{r}-1)^{r}
x^{r}}{(1+q+...+q^{r-1})^{r} }\right) =\sum_{m \geq 0}
(-1)^{m}b_{rm}( x,\, q^{r};{\delta})
\end{align}
where
\begin{align}
 b_{rm}( x,\, q^{r};{\delta})
&=(q^{\delta})^{{r}\binom{m}{2}} \frac{x^{rm}}{[rm]_{q}!}
=(q^{r})^{{\delta}\binom{m}{2}} \frac{x^{rm}}{\alpha_{rm,q}}.
\end{align}

For every $\lambda \in \mathbb{C}$, the function $\cos_{r}(x,\,
q^{r};{\delta})$ is a unique solution of the system
\begin{align}
\begin{cases}
\Lambda_{q\delta}^{-1}D_{q}^{r}u(x) & =\quad  -\lambda^{r} u(x),\\
\quad u(0)&=\quad 1,\; \\ D_{q}^{k}u(0)&=\quad  0,\quad k=1,...,r-1.
\end{cases}
\end{align}

We note $r-q^{\delta}$-sinus of order $( r,l )$ , $l= 1,..., r-1$ by
\begin{align}\label{0002}
\sin_{r,l}(x,\, q^{r};{\delta}) = \sum_{m \geq 0} (-1)^{m}
(q^{\delta})^{{r}\binom{m}{2}} \frac{x^{rm +r-l}}{[rm + r-l]_{q}!}.
\end{align}

Let  $ \mu =e^{i\pi/r} $ and \,$w_{k}= e^{2i\pi(k-1)/r}$,
$k=1,2,...,r$. Since

\begin{equation}
\sum_{k=1}^{r}(w_{k})^{m}=
  \begin{cases}
    r & \text{for integers m  divisible by r}, \\
    0 & \text{for integers m  not divisible by r}.
  \end{cases}
\end{equation}
and expanding the $q$-exponential function in series, we obtain
\begin{equation}
\cos_{r}(x,\, q^{r};{r \delta }) =
\frac{1}{r}\sum_{k=1}^{r}e_{q^{r}}\big(\frac{{\mu
w_{k}{x}}}{{q^{(r-1)/2}}}\; ,\, \delta \big).
\end{equation}
When $r=3,\, \delta = 1$, we obtain the result in \cite{FIT sab}.
\begin{definition}Let  $ x \in \mathbb{R}$ and \,$w_{k}= e^{2i\pi(k-1)/r}$, k=1,2,...,r, a function $f(x)$ is called
$r$ even if
\begin{equation}
f(w_{k}x)=f(x) \qquad k=1,...,r,
\end{equation}
and $r$ odd of $l$ order if
\begin{equation}
f(x)= w_{k}^{l} f(w_{k}x), \qquad k=1,...,r.
\end{equation}

\end{definition}

\begin{proposition}
The functions $cos_{r}$ and $sin_{r,l}$ $(l= 1, ...,r-1 )$are,
respectively,r-even and r-odd of order l. From (\ref{0001}) and
(\ref{0002}) we obtain the following $q$-derivative formulas :
\begin{equation*}
\begin{aligned}
D_{q}^{l}  \cos_{r}(x,\, q^{r};{\delta}) =   - \, q^{-\delta (r-l)}
\sin_{r,l}(q^{\delta} x,\, q^{r};{\delta}),\;\;
D_{q}^{r}\cos_{r}(x,\, q^{r};{\delta}) =  - \cos_{r}(q^{\delta} x,\,
 q^{r};{\delta},),\\
D_{q}^{l-m}\sin_{r,m}(x,\, q^{r};{\delta}) =  \sin_{r,l}(x,\,
 q^{r};{\delta}),\;\;
D_{q}^{r-m}\sin_{r,m}(x,\, q^{r};{\delta}) =  \cos_{r}(x,\,
 q^{r};{\delta}).
\end{aligned}
\end{equation*}

\end{proposition}
\begin{proposition}
The function $\cos_{r}( x,\, q^{r};1 )$ is $r$ even and satisfies,in
particular

\begin{equation}
\cos_{r}( xt,\, q^{r};1) =
(-1)^{n}q^{r(n(n+1)/2)}\frac{1}{x^{rn}}D_{q,t}^{rn}(\cos_{r}(
xtq^{-n},\, q^{r};1)).
\end{equation}
\end{proposition}
\begin{proposition}
Let $x \in \mathbb{R}$ for $n \geq 1$, the function $ b_{rn}(x,\,
q^{r};1)$ verify the following properties
\begin{align}
  b_{0}(x,\, q^{r};1)= 1 ,\;\;  \;  &  b_{rn}(0,\, q^{r};1)=0 \; \,\text{and} \; &  \Lambda_{q}^{-1}D_{q}^{r}b_{rn}(x,\, q^{r};1)=b_{r(n-1)}(x,\,
  q^{r};1).
\end{align}
\begin{equation}\label{0}
 \mid  b_{rn}(x,\, q^{r};{\delta}) \mid \leq \mid  b_{rn}(x,\,
q^{r};1) \mid \leq \frac{q^{-\binom{r}{2}}x^{rn}}{ (rn)!},\; \delta
\geq 1.
\end{equation}
\end{proposition}
\begin{proof}When we put $q = e^{-t},\quad t >0$. The coefficients
$ b_{rn}(x,\, q^{r};1)$ defined by ( \ref{0001}) can be written as
\begin{align}
 b_{rn}(x,\, q^{r};1) &=
 \prod_{j=0}^{n-1}\prod_{i=0}^{r-1}\frac{q^{j}-q^{j+1}}{1-q^{rj+1+i}}
      = \prod_{j=0}^{n-1}\prod_{i=0}^{r-1}\frac{e^{-jt}-e^{-(j+1)t}}{1-e^{-(rj+1+i)t}}\label{000}
 \end{align}
 Preceding like in \cite{koornwinder_92}
; for $j \neq 0$, the following function
 $\displaystyle f_{i}(t) = \frac{e^{-jt}-e^{-(j+1)t}}{1-e^{-(rj+1+i)t}} $ are decreasing in $ ]0,\infty [$, we obtain to something as limite where $t$ tend to $0$ in (\ref{000}).
\end{proof}

When  $|x| \uparrow \infty$, we have
$
 | \cos_{r}(x,\, q^{r};{\delta}) |  \leq q^{- \binom{r}{2}} | \cos_{r}(x) |   \leq   q^{-\binom{r}{2}} \, e^{(r-2)|x| } , \delta \geq
 1, \text{see  \cite{FIT3}}.
$

\subsection{$q^{\delta}$-Product formula }
We set now  the product formula for $q^{\delta}$-cosinus function.
We note by $$P = \cos_{r}(x,\, q^{r};{\delta})\cos_{r}(y,\,
q^{r};{\delta}). $$
\begin{proposition}
Let $x$ and $y$ complex numbers, with $y\neq 0$, we have :\\ $$P =
\sum_{k \geq 0} \frac{(q^{\delta})^{{r}{k}^{2}}}{(1-q)^{rk}[r
k]_{q}!}(-1)^{k} q^{-\binom{rk}{2}}\big(\frac{x}{y} \big)^{rk}
\sum_{s=0}^{rk} (-1)^{s} q^{\binom{s}{2}}\left[\begin{matrix} {rk}
\\ {s}\end{matrix}\right]_{q} \Lambda_{q^{\delta}}^{-k} \cos_{r}(y
q^{rk-s} ,\, q^{r};{\delta}). $$
\end{proposition}
\begin{proof}
For $y\neq
 0$
$$ P = \sum_{k \geq 0} \frac{(q^{\delta})^{{r}{k}^{2}}}{[r
k]_{q}!}\big(\frac{x}{y} \big)^{rk}\sum_{n \geq 0}
(-1)^{n}\frac{(q^{r\delta})^{\binom{n}{2}}}{[r(n-k)]_{q}!}(q^{\delta})^{-rnk}
y^{rn}. $$
 Moreover, if we use the previous relation
$$\frac{[rn]_{q}!}{[r(n-k)]_{q}!}(1-q)^{rk} = (-1)^{rk}
q^{-\binom{rk}{2}+r^{2}n k} \sum_{s=0}^{rk} (-1)^{s}
q^{\binom{s}{2}}\left[\begin{matrix} {rk} \\
{s}\end{matrix}\right]_{q} q^{-rns},$$ we obtain that $$ P = \sum_{k
\geq 0} \frac{(q^{\delta})^{{r}{k}^{2}}}{(1-q)^{rk}[r
k]_{q}!}(-1)^{k} q^{-\binom{rk}{2}}\big(\frac{x}{y} \big)^{rk}
\sum_{s=0}^{rk} (-1)^{s} q^{\binom{s}{2}}\left[\begin{matrix} {rk}
\\ {s}\end{matrix}\right]_{q}  \cos_{r}(yq^{k(r-\delta)}q^{-s} ,\,
q^{r};{\delta}).$$
\end{proof}

\section{The $q$-Bessel Operator of $r$-order }
We suppose now that the components of the vector  $\alpha = (\alpha
_1 ,...,\alpha _{r - 1} )$ where $\alpha_{k}$ is a reel number
satisfy  $\alpha _k \ge - 1 + \frac{k}{r}, \quad k = 1,...,r - 1$
and $\delta
>0$.

The $q$-Bessel operator of $r$-order is defined by
\begin{equation}\label{ui}
  B_{r,\delta} u = \Lambda_{q^{\delta}}^{-1}\big(\frac{1}{x^{r-1}}\prod_{i=1}^{r-1}\big(q^{r\alpha _i + 1} x D_{q} + ({ r \alpha _{i} + 1})_{q} \;  \big)
   D_{q}u \big).
\end{equation}
\begin{remark}
  For  $r = 2$, we obtain the $q$-Bessel operator $B_{2,\delta} $ of
  the second order studied in \cite{FIT3} for $\delta = 1$
\begin{equation}\label{b2}
  B_{2,\delta} u = \Lambda_{q^{\delta}}^{-1}\big(q^{2\alpha  + 1} D_{q}^{2}u +  \frac{(2\alpha  + 1)_{q}}{x}D_{q}u
  \big).
\end{equation}
and for $r = 3$, $\alpha _1 = -2/3$, $\alpha _2=\nu-1/3$, we obtain
the operator $B_{3,\delta} $ studied in \cite{FIT sab}
\begin{equation}
B_{3,\delta} u = \Lambda_{q^{\delta}}^{-1}\big( q^{3\nu} D_{q}^{3}u
+ \frac{1}{q} \frac{(3\,\nu)_{q}}{x}\,D_{q}^{2}u - \frac{1}{q}
\frac{(3\,\nu)_{q}}{x^2}\,D_{q}u\big).
 \label{equ1}
\end{equation}

\begin{proposition}
For $\lambda$ being a complex number, the function
${{j}}_{\alpha}(\lambda x,q^{r},\delta)$  is a solution of the
$q$-problem
\begin{equation}
\begin{aligned}
 B_{r,\delta}u(x)& =  - \lambda^{r} u(x)  \\ u(0) = 1, &\;
D_{q}^{k}u(0)  = 0,\quad k = 1,...,r - 1\, .
\end{aligned} \label{equ2q}
\end{equation}

\end{proposition}

\begin{equation}
\begin{aligned}
{{j}}_{\alpha}(\lambda x,q^{r},\delta)=& \sum_{n=0}^{\infty}
(-1)^{n}b_{rn,\alpha}(x,q^{r},\delta) \lambda^{rn},\\
b_{rn,\alpha}(x,q^{r},\delta)=&\frac{(q^{r})^{\delta
\binom{n}{2}}x^{rn}}{((r)_{q})^{rn}(1)_{n}^{q^{r}}{\prod
}(\alpha_{i} +1 )_{n}^{q^{r}}} = \frac{(q^{r})^{\delta
\binom{n}{2}}x^{rn}}{\alpha_{rn,\alpha,q}}, \\
\alpha_{rn,\alpha,q} =&
(1+q+...+q^{r-1})^{rn}\,[n]_{q^{r}}!\,\prod_{i=1}^{r-1}
\frac{\Gamma_{q^{r}}(\alpha_{i} + n+1)}{\Gamma_{q^{r}}(\alpha_{i}
+1)}.
\end{aligned}\label{j}
\end{equation}
\end{remark}
\begin{align}
{{j}}_{\alpha}( x,q^{r},\delta)=
{_{0}\phi_{r-1}^{\delta}}\left(\begin{matrix} - ,
\\ (q^{r})^{\alpha_{1}+1},...,(q^{r})^{\alpha_{r-1} +1}
\end{matrix}\bigg|q^{r};- \frac{(q^{r}-1)^{r}
x^{r}}{(1+q+...+q^{r-1})^{r} }\right).
\end{align}
For $\delta = r$, we obtain the $q$-hypergeometric function
$_{0}\phi_{r-1}$.

\subsection{ Increase of $b_{rn,\alpha}(x,q^{r},\delta)$ :}Let  now $\mid \alpha \mid =
\alpha_{1} + ....+\alpha_{r-1}=\alpha_{0} + ....+\alpha_{r-1}$ with
$ \alpha_{0} = 0$

\begin{equation}\label{12}
b_{rn,\alpha}(1,\, q^{r};{\delta}) \leq b_{rn,\alpha}(1,q^{r},1) =
\frac{(q^{r})^{
\binom{n}{2}}}{((r)_{q})^{rn}(1)_{n}^{q^{r}}{\prod}(\alpha_{i} +1
)_{n}^{q^{r}}}, \quad \delta \geq 1
\end{equation} the right term can be written by
\begin{equation}\label{122}
\frac{\big((q^{r})^{-\mid \alpha \mid /r}\big)^{n}}{((r)_{q})^{rn}}
\prod_{j=0}^{n-1} \prod_{i=0}^{r-1} \frac{(q^{r})^{(\alpha_{i} +
j)/r} - (q^{r})^{1+(\alpha_{i} + j)/r} }{1 - (q^{r})^{1+(\alpha_{i}
+ j)}}.
\end{equation}

Now, by \cite{koornwinder_90} ,\; lemma A.{1} and
\cite{koornwinder_92} ,\; proposition A.{2}, we see that the general
terms of product increases to $( j + \alpha_{i} + 1)^{-1}$ if $q
\uparrow 1$. Using Stirling's formula, we find that, for some
constant $C$.
\begin{align*}\label{33}
 b_{rn,\alpha}(1,\, q^{r};{\delta}) \leq   \frac{\big(  (q^{r})^{- |\alpha|/r} \big)^{n}}{((r)_{q})^{rn} \prod (\alpha_{i}+1)_{n}
 }
 \leq  \; C \; \big(  (q^{r})^{- |\alpha|/r} \big)^{n} \bigg(\frac{e}{n (r)_{q}}\bigg)^{rn +
 |\alpha|},
\end{align*}
this inequality generalizes the inequality in \cite{FIT sab}.

Use the functional equation of $q$-Gamma, we obtain the following
proposition :
\begin{proposition}
For $\alpha _i \ge - 1 + \frac{i}{r},\;  i = 1,...,r - 1$, and
n=0,1,2..., we have
\begin{equation}
D_{q}{{j}}_{\alpha}( . ,q^{r},\delta)(x)= -
\bigg(\frac{x}{(r)_{q}}\bigg)^{r-1} \frac{1}{\prod (\alpha_{i} +1
)_{q^{r}}} {{j}}_{\alpha+1}( q^{\delta}x ,q^{r},\delta),
\end{equation}

and
\begin{equation}
\{\frac{1}{x^{r-1}}D_{q}\}^{n}{{j}}_{\alpha}( x ,q^{r},\delta)=
\bigg(\big(\frac{1}{(r)_{q}}\big)^{r-1}\bigg)^{n} \frac{(-1)^{n}
(q^{ \delta})^{\binom{n}{2}}}{\prod (\alpha_{i} +1 )_{n}^{q^{r}}}
{{j}}_{\alpha+n}( q^{n \delta}x ,q^{r},\delta).
\end{equation}
\end{proposition}
 By the $q$-duplication formula of  $\Gamma_{q}$ (\ref{001}), we have, in particular

\begin{equation}\label{0}
  {{j}}_{(-1/r,-2/r,...,-(r-1)/r)}( x ,q^{r},\delta) =\cos_{r}(x,\,
  q^{r};{\delta}).
\end{equation}

\section{$q$-Integral Representations}

In this section, we give two $q$-integral representations of the
${{q-j}}_{\alpha}$ function (\ref{j}) involving the $q$-Jackson
integral. We denote by $W_{\alpha}$ the function
\begin{align}
W_{\alpha}(t_{1},..., t_{r-1}; q^{r}) =& \prod_{i=1}^{r-1}
\frac{(t_{i}^{r}q^{r};\,q^{r})_{\infty}}{(t_{i}^{r}q^{\alpha_{i} -
\frac{i}{r}+1};\,q^{r})_{\infty}}t_{i}^{i-1} =\prod_{i=1}^{r-1}
(t_{i}^{r}q^{r};\,q^{r})_{\alpha_{i} - \frac{i}{r}}t_{i}^{i-1}\\
=& \prod_{i=1}^{r-1}(1-q^{r}t_{i}^{r})_{q^{r}}^{\alpha_{i} -
\frac{i}{r}}t_{i}^{i-1}
\end{align}\label{eq4q}
which tends to $\prod(1- t_{i}^{r})^{\alpha_{i} -
\frac{i}{r}}\;t_{i}^{i-1}$ as $q \longrightarrow 1^{-}$\; .

\subsection{$q$-Mehler Type}

\begin{theorem} \label{thm116}

For $\alpha _i \ge - 1 + \frac{i}{r},\;  i = 1,...,r - 1$, the
function ${{j}}_{\alpha}$ has the following $q$-integral
representation of Mehler type
\begin{equation}
{{j}}_{\alpha}( z ,q^{r},\delta)=C_{r,\alpha}
\int_{0}^{1}...\int_{0}^{1} W_{\alpha}(t_{1},..., t_{r-1}; q^{r}) \;
\cos_{r}(zt_{1},..., t_{r-1};q^{r},\delta)\,
d_{q}t_{1}...d_{q}t_{r-1},
\end{equation}

where
\begin{equation}
C_{r,\alpha}=((r)_{q})^{r-1}.\prod_{i=1}^{r-1}
\frac{\Gamma_{q^{r}}(\alpha_{i}
+1)}{\Gamma_{q^{r}}(\frac{i}{r})\Gamma_{q^{r}}(\alpha_{i} -
\frac{i}{r}+1 )}.
\end{equation}
\end{theorem} \label{thm16}
\begin{proof}
This formula can be proved by expanding $\cos_{r}(zt,q^{r};\delta)$
in a series of power of $t$ and then $i$ grating, there arise
$q$-integrals of the form
\begin{equation}
\int_{0}^{1} t_{i}^{rm}(1-q^{r}t_{i}^{r})_{q^{r}}^{\alpha_{i} -
\frac{i}{r}}t_{i}^{i-1}\,
d_{q}t_{i}=\frac{\Gamma_{q^{r}}(m+\frac{i}{r})\Gamma_{q^{r}}(\alpha_{i}
- \frac{i}{r}+1 )}{(r)_{q} \Gamma_{q^{r}}(\alpha_{i} +m+1)}.
\end{equation}

Basis of the  $q$-duplication formula for the $\Gamma_{q}$ function
(\ref{001}), the  formula is proved.
\end{proof}
\begin{proposition}
For $\alpha _i \ge - 1 + \frac{i}{r},\;  i = 1,...,r - 1$, and
n=0,1,2..., we have
\begin{equation}
\bigg|D_{q}^{n}\big[{{j}}_{\alpha}( x ,q^{r},\delta) \big]\bigg|\leq
\prod_{i=1}^{r-1} \frac{\Gamma_{q^{r}}(\alpha_{i}
+1)\Gamma_{q^{r}}(\frac{n+i}{r})}{\Gamma_{q^{r}}(\frac{i}{r})\Gamma_{q^{r}}(\alpha_{i}
+1+\frac{n}{r})}\bigg| \big[ D_{q,x}^{n} \cos_{r}(x,\,
q^{r};{\delta})\big]\bigg|,
\end{equation}
in particular
\begin{equation}
\big|{{j}}_{\alpha}( x ,q^{r},\delta)\big|\leq q^{-\binom{r}{2}} \,
e^{(r-2)|x| }.
\end{equation}
\end{proposition}

\subsection{$q$-Sonine Type}

\begin{theorem} \label{thm116}

For $\alpha _i \ge - 1 + \frac{i}{r},\;  i = 1,...,r - 1$ and $p_{i}
\geq 1$,the function ${{j}}_{\alpha + p}$ has the following
$q$-integral representation of Sonine type
\begin{equation}
{{j}}_{\alpha+p}( z ,q^{r},\delta)=D_{r,\alpha,p}\,
\int_{0}^{1}...\int_{0}^{1} \; V_{p}(t_{1},..., t_{r-1};
q^{r})\;{{j}}_{\alpha}(z t_{1},..., t_{r-1},q^{r};\delta)\;
  \,
d_{q}t_{1}...d_{q}t_{r-1},
\end{equation}
where
\begin{equation}
D_{r,\alpha,p}=((r)_{q})^{r-1}.\prod_{i=1}^{r-1}
\frac{\Gamma_{q^{r}}(\alpha_{i}+ p_{i}
+1)}{\Gamma_{q^{r}}(p_{i})\Gamma_{q^{r}}(\alpha_{i}+1 )}
\end{equation}
\begin{equation}
 V_{p}(t_{1},..., t_{r-1};
q^{r})=  \prod_{i=1}^{r-1}(1-q^{r}t_{i}^{r})_{q^{r}}^{p_{i}}\;
t_{i}^{r(\alpha_{i} - \frac{i}{r}+1 )}\; t_{i}^{i-1}\; .
\end{equation}

\end{theorem} \label{thm17}
\begin{proof}

This formula can be proved by expanding ${{j}}_{\alpha}$ in a series
of power of $t_{i}$ , there arise $q$-integrals of the form
$$
\int_{0}^{1} t_{i}^{r(\alpha_{i} -
\frac{i}{r}+1+m)}(1-q^{r}t_{i}^{r})_{q^{r}}^{p_{i} - 1}t_{i}^{i-1}\,
d_{q}t_{i}=\frac{\Gamma_{q^{r}}(\alpha_{i} +m+1)\Gamma_{q^{r}}(p_{i}
)}{(r)_{q} \Gamma_{q^{r}}(m +\alpha_{i}+ p_{i} +1)}.
$$
\end{proof}

\section{$q$-Fourier Transform }
Notation. - Some $q$-functional spaces will be used to establish our
result. We putting
\begin{itemize}
  \item $\mathbb{R}_{q} = \{\pm q^{k}, \; k\in \mathbb{Z}    \} \cup \{ 0
  \}$,
 \quad $\mathbb{R}_{q}^{*} = \{\pm q^{k}, \; k\in \mathbb{Z}
 \},$

  \item $\mathbb{R}_{q,+} = \{ q^{k}, \; k\in \mathbb{Z}    \} \cup \{ 0
  \}$,
 \quad  $\mathbb{R}_{q,+}^{*} = \{ q^{k}, \; k\in \mathbb{Z}
  \}$.
\end{itemize}

- We design by $\mathcal{E}_{*,q}(\mathbb{R})$ (resp
${\mathcal{E}_{*,\,q}}(\mathbb{R}_{q})$) the space of $r$-even
functions defined on $\mathbb{R}$ (resp $\mathbb{R}_{q}$) infinitely
$q$-derivative, and by ${\mathcal{D}_{*,\,q}}(\mathbb{R})$ (resp
${\mathcal{D}_{*,\,q}}(\mathbb{R}_{q})$)
 the space of $r$-even functions defined on $\mathbb{R}$ (resp $\mathbb{R}_{q}$) infinitely
$q$-derivative with compact support.

In this section we introduce the space
$\mathfrak{L}_{\alpha,\,q^{\delta}}^{1}(\mathbb{R}_{q,+},\,d_{q}x)$
of functions $f$ satisfying $$\int_{0}^{\infty} |f(x){{j}}_{\alpha}(
\lambda x,\, q^{r};{\delta})|\,d_{q}x < \infty,\qquad  \lambda \in
\mathbb{R}_{q}.$$
\begin{definition}
The Fourier  transform related with $ B_{r,\delta}$ of $f \in
\mathfrak{L}_{\alpha,\,q^{\delta}}^{1}(\mathbb{R}_{q,+},\,d_{q}x)$
is the function $\mathcal{F}_{q^{\delta}}(f)$ defined by

\begin{equation}\label{tf}
 \mathcal{F}_{q^{\delta}}(f)(\lambda) =\int_{0}^{ \infty} \, f(t) \, {{j}}_{\alpha}(\lambda t,\, q^{r};{\delta})\, d_{q}t,\quad \lambda \in
 \mathbb{R}_{q}.
\end{equation}
We define also the Fourier  transform  $ \mathcal{F}_{0,q^{\delta}}$
by
\begin{equation}
 \mathcal{F}_{0,q^{\delta}}(f)(\lambda) =\int_{0}^{ \infty} \, f(t) \, {\cos}_{r}(\lambda t,\, q^{r};{\delta})\, d_{q}t,\quad \lambda \in
 \mathbb{R}_{q}.
\end{equation}\label{thm7.1}
\end{definition}

\section{$q$-Translation and $q$-Convolution}

In this section we study the generalized translation operator
associated with the operator $B_{r,\delta}$ . We give the following
definition related to $\Lambda_{q\delta}^{-1}D_{q}^{r}$.

\begin{definition}
The translation operator $\tau_{x,\,q^{\delta}}$, $x \in \mathbb{R}$
(resp $\mathbb{R}_{q}$) associated with the $r$-order derivative
operator $\Lambda_{q\delta}^{-1}D_{q}^{r}$ is defined for $f$ in
$\mathcal{E}_{*,q}(\mathbb{R})$  (resp
${\mathcal{E}_{*,\,q}}(\mathbb{R}_{q})$) and  $y\in \mathbb{R}$
(resp $\mathbb{R}_{q}$) by

\begin{equation}
\tau_{x,\,q^{\delta}}(f) (y) = \sum_{n = 0}^{\infty}  b_{rn}(y,\,
q^{r};{\delta})\, (\Lambda_{q\delta}^{-1}D_{q}^{r})^{(n)}\,f(x),
\end{equation}
the functions $ b_{rn}(y,\, q^{r};{\delta})$ are given by
(\ref{0001}).
\end{definition}

We have the product formula $$ \cos_{r}(\lambda x,\,
q^{r};{\delta}).\cos_{r}(\lambda y,\, q^{r};{\delta}) =
\tau_{x,\,q^{\delta}} \cos_{r}(\lambda y,\, q^{r};{\delta}) =
\tau_{y,\,q^{\delta}} \cos_{r}(\lambda x,\, q^{r};{\delta}). $$

\begin{proposition}\label{prop8.1}
The operators $\tau_{x,\,q^{\delta}}$ satisfy :
\begin{enumerate}
\item  For $x \in \mathbb{R}$, $\tau_{x,\,q^{\delta}}$ belong in $\mathcal{L} ( \mathcal{E}_{*,q}(\mathbb{R}),\,
\mathcal{E}_{*,q}(\mathbb{R})$).
\item  The map $x \longrightarrow \tau_{x,\,q^{\delta}}$ is infinitely $q$-derivative, $r$-even.
\end{enumerate}
\end{proposition}

\begin{lemma}\label{001.222}
For $f \in D_{*,q}(\mathbb{R})$, $n \in \mathbb{N}$, we have :
\begin{equation*}
\big(\Lambda_{q^{\delta}}^{-1}D_{q}^{r}\big)^{n}f(x) =
\frac{q^{-\binom{rn}{2}} }{(1-q)^{rn}(q^{-\delta n })^{rn}
b_{rn}(x,\,
q^{r};{\delta})}\sum_{k=0}^{rn}\frac{(-1)^{k}q^{\binom{rn-k}{2}}}{[rn-k]_{q}![k]_{q}!}
\Lambda_{q^{\delta}}^{-n} f(q^{k}x).
\end{equation*}
\end{lemma}
\begin{proof}
For $\delta > 0$, by \cite{koornwinder_1999} and (\ref{intr10pp})
\begin{equation*}
D_{q,x}^{rn}f(x) =
\frac{q^{-\binom{rn}{2}}}{(1-q)^{rn}x^{rn}}\sum_{k=0}^{rn}(-1)^{k}
\left[\begin{matrix}{rn} \\ {k}\end{matrix}\right]_{q}
q^{\binom{rn-k}{2}}  f(q^{k}x),
\end{equation*}
\begin{equation*}
\big(\Lambda_{q^{\delta}}^{-1}D_{q}^{r}\big)^{n} = \big(
(q^{\delta})^{r}
\big)^{-\binom{n}{2}}\Lambda_{q^{\delta}}^{-n}D_{q}^{rn},
\end{equation*}
\begin{equation*}
\big(\Lambda_{q^{\delta}}^{-1}D_{q}^{r}\big)^{n}f(x) =
\frac{q^{-\binom{rn}{2}} \big( (q^{\delta})^{r}
\big)^{-\binom{n}{2}}}{(1-q)^{rn}(q^{-\delta n
}x)^{rn}}\sum_{k=0}^{rn}(-1)^{k} \left[\begin{matrix} {rn} \\
{k}\end{matrix}\right]_{q} q^{\binom{rn-k}{2}}
\Lambda_{q^{\delta}}^{-n} f(q^{k}x).
\end{equation*}
\end{proof}
We obtain for $\delta > 0$
\begin{equation*}
\tau_{y,q^{\delta}}f(x) = \sum_{n=0}^{\infty}\frac{b_{rn}(1,\,
q^{r};{\delta})}{(1-q)^{rn}(q^{-\delta n
})^{rn}}\bigg(\frac{y}{x}\bigg)^{rn}q^{-\binom{rn}{2}}
\sum_{k=0}^{rn}\,(-1)^{k}\left[\begin{matrix} {rn} \\
{k}\end{matrix}\right]_{q} q^{\binom{rn-k}{2}}
\Lambda_{q^{\delta}}^{-n} f(q^{k}x).
\end{equation*}

\begin{proposition}
For $f \in \mathcal{D}_{*,\,q}(\mathbb{R}_{q})$ we have :
\begin{equation*}
\mathcal{F}_{0,\,q^{\delta}} ( {^{t}  \tau_{x,\,q^{\delta}}} f
)(\lambda) = \cos_{r}(\lambda x,\, q^{3};{\delta})
\mathcal{F}_{0,\,q^{\delta}} ( f )(\lambda).
\end{equation*}
The convolution product of two functions $ f$ and $g$ of \;
$\mathcal{D}_{*,\,q}(\mathbb{R}_{q})$ is defined  by :
\begin{equation*}
f \star_{q^{\delta}} \,g (x) = \int_{0}^{\infty}\, {^{t}
\tau_{x,\,q^{\delta}}} f(y) g(y) \,d_{q}y = \int_{0}^{\infty}\,f(y)
{ \tau_{x,\,q^{\delta}}}  g(y) \,d_{q}y.
\end{equation*}
in $\mathcal{D}_{*,\,q}(\mathbb{R}_{q})$ and we have :
$
\mathcal{F}_{0,\,q^{\delta}} ( f \star_{q^{\delta}}\, g )(\lambda) =
\mathcal{F}_{0,\,q^{\delta}} (f)(\lambda) .\,
\mathcal{F}_{0,\,q^{\delta}} (g)(\lambda).
$
\end{proposition}

These  previous properties can be extended for the operator
$B_{r,\delta}$ and  suggest the following  definition

\begin{definition}
\text{We call generalized translation operators  associated with }
$B_{r,\delta}$\, the operators
$\mathrm{T}_{x,\,q^{\delta}}^{\alpha}$, $x \in \mathbb{R}$ (resp
$\mathbb{R}_{q}$),\,defined on $\mathcal{E}_{*,\,q}(\mathbb{R})$
(resp ${\mathcal{E}_{*,\,q}}(\mathbb{R}_{q})$) by :
\begin{equation}
\mathrm{T}_{x,\,q^{\delta}}^{\alpha}(f)(y) = \sum_{n = 0}^{\infty}
b_{rn,\,\alpha }(y,\, q^{r};{\delta})\,
B_{r,\delta}^{n}\,(f)(y),\qquad y \in  \mathbb{R}\;(\text{resp}\;
\mathbb{R}_{q}),
\end{equation}
where the functions $b_{rn,\alpha}(y,q^{r},\delta)$ is given by
(\ref{j}).
\end{definition}

We summarize the properties of
$\mathrm{T}_{x,\,q^{\delta}}^{\alpha}$ in this proposition

\begin{proposition}
The operators  $\mathrm{T}_{x,\,q^{\delta}}^{\alpha}$ satisfy :
\begin{enumerate}
 \item  \text{For} $x \in \mathbb{R},\, \mathrm{T}_{x,\,q^{\delta}}^{\alpha}$ \text{in}\,  $\mathcal{L} ( \mathcal{E}_{*,q}(\mathbb{R}),\,  \mathcal{E}_{*,q}(\mathbb{R}))$
 \item  The map $x \longrightarrow \mathrm{T}_{x,\,q^{\delta}}^{\alpha}$ are infinitely $q$-derivative and $r$-even.
 \item  For all functions $ f$ in $\mathcal{E}_{*,q}(\mathbb{R})$ :\\
   -- \;$ \mathrm{T}_{x,\,q^{\delta}}^{\alpha}f (y) = \mathrm{T}_{y,\,q^{\delta}}^{\alpha}f (x);$\\
   -- \;$ \mathrm{T}_{0,\,q^{\delta}}^{\alpha}f (y) = f (y).$
\item  For given $f$ in $\mathcal{E}_{*,q}(\mathbb{R})$, we put : $ u(x,\;y) =  \mathrm{T}_{x,\,q^{\delta}}^{\alpha} \,  f ( y ). $
\end{enumerate}

Then the function $ u$ is  solution of the Cauchy problem :
\begin{equation*}
(\mathbb{I})\qquad
  \begin{cases}
  \begin{aligned}
     &B_{x,r,\delta} u(x,\,y) = B_{y,r,\delta} u(x,\,y) , \\
    &u(x,\,0) = f(x);D_{q,\,y} u(x,\,0) = 0 ;\\& D_{q,y}^{k} u(x,\,0) = 0 \, k = 0.1.2...r-1 .
\end{aligned}
  \end{cases}
\end{equation*}
\end{proposition}

\begin{equation}
\mathrm{T}_{x,\,q^{\delta}}^{\alpha}{{j}}_{\alpha}(\lambda y
,q^{r},\delta)= {{j}}_{\alpha}(\lambda x,q^{r},\delta)
{{j}}_{\alpha}(\lambda y,q^{r},\delta) =
\mathrm{T}_{y,\,q^{\delta}}^{\alpha}{{j}}_{\alpha}(\lambda
x,q^{r},\delta).
\end{equation}
Now we are able to define the convolution product related to the
operator $B_{r,\delta}$.
\begin{definition}
The convolution  product associated with $B_{r,\delta}$  of two
functions $f$ and  $g$ in $\mathcal{D}_{*,\,q}(\mathbb{R}_{q})$ is
the function $f \star_{\alpha,\,q^{\delta}}\, g$ defined by :
\begin{equation}
f \star_{\alpha,\,q^{\delta}}\, g (y) =  \int_{0}^{\infty}\,f(x)
\mathrm{T}_{y,\,q^{\delta}}^{\alpha}g(x) \, d_{q}x =
\int_{0}^{\infty}\,^{t}\mathrm{T}_{y,\,q^{\delta}}^{\alpha}f(x)
\,g(x) \, d_{q}x.
\end{equation}
\end{definition}

\section{ Higher-order $q$-Bessel Heat Polynomials }

We recall that the function  $e_{q^{r}}(-z^{r}t ){{j}}_{\alpha}( x z
;\,q^{r} ;\, \delta)$ is analytic in $z^{r}$. We thus have, for
$t\in \mathbb{R}$ and $\delta \geq 1$,
\begin{align}
& e_{q^{r}}(-z^{r}t ){{j}}_{\alpha}(  x z  ;\,q^{r} ;\, \delta) =
\sum_{n=0}^{\infty}(-1)^{n}\frac{z^{rn}}{\alpha_{rn,\alpha,q}}p_{n}^{\alpha}(x,\,t,\,q^{r}
;\, \delta)\\ & p_{n}^{\alpha}(x,\,t,\,q^{r} ;\, \delta)=
\sum_{k=0}^{n}(q^{r})^{\delta
\binom{n-k}{2}}\frac{(x^{r})^{n-k}t^{k}}{[k]_{q^{r}}!}\frac{{\alpha_{rn,\alpha,q}}}{{\alpha_{r(n-k),\alpha,q}}}\\
=& \frac{\prod(\alpha_{i} +1
)_{n}^{q^{r}}}{(1+q+...+q^{r-1})^{-rn}}t^{n}\sum_{k=0}^{\infty}\frac{(-1)^{k}(-n)_{k}^{q^{r}}
(q^{r})^{(\delta-1)\binom{k}{2}}(q^{r})^{n k}(x^{r})^{k}t^{-k}
}{\prod (\alpha_{i} +1 )_{k}^{q^{r}} (1+q+...+q^{r-1})^{rk}
[k]_{q^{r}}!}\\ =&
\frac{{\alpha_{rn,\alpha,q}}}{[n]_{q^{r}}!}\,t^{n}\,_{1}\phi_{r-1}^{\delta-1}\left(\begin{matrix}
(q^{r})^{-n}, \\ (q^{r})^{\alpha_{1}+1},...,(q^{r})^{\alpha_{r-1}
+1}
\end{matrix}\bigg|q^{r};\frac{(q^{r} -1)^{r-1}(-
x^{r}(q^{r})^{n})}{(1+q+...+q^{r-1})^{r} t}\right).\label{pnqq}
\end{align}
\section{$q$-Heat Equation}
We give an applications of the Fourier  transform related with $
B_{r,\delta}$. We begin by recalling that
\begin{align}\label{Iq}
\int_{0}^{\infty}e_{q^{r}}(-c x^{r}) \, ( c^{n} x^{rn})x^{r
\alpha_{k} + (r-1)}\, d_{q}x =
\frac{(q^{r})^{-n(\alpha_{k}+1)-\binom{n}{2}}(\alpha_{k}+1)_{n}^{q^{r}}
\mathbf{I}(\alpha_{k} +
1,q^{r})}{c^{\alpha_{k}+1}(1+q+...+q^{r-1})}\\
 \mathbf{I}(\alpha_{k} + 1;\,q^{r}) = \int_{0}^{\infty}e_{q^{r}}(-x) \, x^{\alpha_{k} }\,
 d_{q^{r}}x\;\; \text{ and } \; \mathbf{H}_{q^{r}}(\alpha_{k} + 1) =
\frac{\mathbf{I}(\alpha_{k} + 1;\,q^{r})}{\Gamma_{q^{r}}(\alpha_{k}
+ 1)}.
\end{align}
We note by $ d\eta_{q,\alpha_{k}}(y) = \frac{y^{r\alpha_{k} + (r-1)
}}{(1+q+...+q^{r-1})^{\alpha_{k}}\Gamma_{q^{r}}(\alpha_{k} + 1)}\,
d_{q}y  $ and we define for $\delta > 1$ the fundamental solution
$\mathcal{K}_{\alpha_{k}}(x,t,\, q^{r};{\delta})$ by
\begin{align*}
&\mathcal{K}_{\alpha_{k}}(x,t,\, q^{r};{\delta}) =
\int_{0}^{\infty}e_{q^{r}}(-t y^{r}){{j}}_{\alpha}( x y ,\,
q^{r};{\delta})\, d\eta_{q,\alpha_{k}}(y), \\ =\, &
\frac{\mathbf{H}_{q^{r}}(\alpha_{k} + 1
)}{(t(1+q+...+q^{r-1}))^{\alpha_{k} + 1}} \times
\sum_{n=0}^{\infty}\frac{(-1)^{n}(q^{\delta})^{r\binom{n}{2}}\times
(q^{r})^{-(\alpha_{k} +
1)n-\binom{n}{2}}x^{rn}t^{-n}}{(1+q+...+q^{r-1})^{rn}[n]_{q^{r}}!
{\prod_{i\neq k} (\alpha_{i} + 1)_{n}^{q^{r}}}},\\
 =&\frac{\mathbf{H}_{q^{r}}(\alpha_{k}+1)}{(t(1+q+...+q^{r-1}))^{\alpha_{k}+1}}\times
\\ & _{0}\phi_{r-2}^{\delta - 1} \left(\begin{matrix}
-,
\\ (q^{r})^{\alpha_{1}+1},..,(q^{r})^{\alpha_{k-1}+1},(q^{r})^{\alpha_{k+1}+1},...,(q^{r})^{\alpha_{r-1} +1}
\end{matrix}\bigg|q^{r};  (\frac{-x^{r}(q^{r})^{-(\alpha_{k}+1)}(q^{r} -
1)^{r-1}}{(1+q+...+q^{r-1})^{r}t})\right).
\end{align*}
For $\delta = r$, we obtain the basic hypergeometric series\ .

We consider the $q$-problem for $ t, x \geq 0$

\begin{equation}
(\mathrm{\,II\,})
\begin{cases}
\begin{aligned}
B_{r,\delta}u(x,\,t)& = D_{q^{r},\,t} u(x,\,t)\\
D_{q}^{k}u(0,\,t)&= 0,\quad k = 1,...,r - 1  \\ u(w_{k}\,x,\,t)&=
u(x,\,t) ,\quad k = 1,...,r - 1  \\ u(x,0)&=f(x).
\end{aligned}
\end{cases}
\end{equation}

\begin{theorem}Let $f \in
\mathfrak{L}_{\alpha,\,q^{\delta}}^{1}(\mathbb{R}_{q,+},\,d_{q}x)$,
the function
\begin{align}
u(x,\,t) =
\int_{0}^{\infty}\mathrm{T}_{y,\,q^{\delta}}^{\alpha}\mathcal{K}_{\alpha_{k}}(x,t,\,
q^{r};{\delta})\,f(y)\,d_{q}y = \big( f \star_{\alpha,q^{\delta}}\,
\mathcal{K}_{\alpha_{k}}(. \,,\,t,\, q^{r};{\delta}) \big)(x),
\end{align}
is a solution of the equation $( II )$ for  $\alpha _k \ge - 1 +
\frac{k}{r},  k = 1,...,r - 1,\, t, x \in \mathbb{R}_{q,+}$.
\end{theorem}

\section{ Analytic Cauchy Problem Related to The $r$-order $q$-Bessel operator $B_{r,\delta}$ }

We say that a function $u(x, t)$ in ${\mathcal{H}}_{\alpha}([0, a]
\times [0, \sigma])$ if
\begin{equation}\label{11}
  B_{r,\delta}\,u(x, t) = D_{q^{r},t}\,u(x, t)
\end{equation}

 $x\in[0, a]$
 and $t\in[0, \sigma]$. The diffusion polynomials $p_{n}^{\alpha}(x, t)$ satisfy the
 $q$-equation(\ref{11}). Hence we expect to obtain infinite series expansions
$ u(x, t) = \sum_{m = 0}^{\infty}a_{m}p_{m}^{\alpha}(x,\,t,\,q^{r}
;\, \delta))$ with possible convergence in a strip $|t| < \sigma$.

Let $\delta \geq 1$, we note
\begin{align*}
{\mathcal{R}}_{\alpha,q}^{\delta}(x) =&  \sum_{n =
0}^{\infty}\frac{(q^{r})^{(\delta-1)\binom{n}{2}}
x^{rn}}{(1+q+...+q^{r-1})^{rn}\prod(\alpha_{i} +1 )_{n}^{q^{r}}}\\
=&  _{1}\varphi_{r-1}^{\delta - 1}\left(\begin{matrix} (q^{r})^1
\\ (q^{r})^{\alpha_{1}+1},...,(q^{r})^{\alpha_{r-1}
+1} \end{matrix}\bigg|q^{r} ; \frac{(q^{r} -
1)^{r-1}x^r}{(1+q+..+q^{r-1})^{r}}\right)\,.
\end{align*}

\begin{lemma} \label{Lem1.3}
Let $s > 0$ and $\delta \geq 1$
\[\frac{p_{n}^{\alpha}(|x|, |t|, q^{r},\delta)}{\alpha_{rn,\alpha,q}} \leq \frac{s^n}{[n]_{q^{r}}!}\,
(1 + \frac{|t|}{s})^n{\mathcal{R}}_{\alpha,q}^{\delta}
\big(\frac{|x|}{s^{1/r}}\big).\]
\end{lemma}

\begin{proof} We have
\begin{align*}
\frac{p_{n}^{\alpha}(|x|, |t|, q^{r},\delta)}{\alpha_{rn,\alpha,q}}
& \leq \frac{s^n}{[n]_{q^{r}}!}\, \sum_{k = 0}^{\infty}
\left[\begin{matrix}{n} \\
{k}\end{matrix}\right]_{q^{r}}\left(\frac{|t|}{s}\right)^{n - k}\,
\frac{(q^{r})^{\delta
\binom{k}{2}}\,\frac{|x|^{rk}}{s^{k}}}{((r)_{q})^{rk}\prod(\alpha_{i}
+1 )_{n}^{q^{r}}} \\
 & \leq  {\mathcal{R}}_{\alpha,q}^{\delta}(\frac{|x|}{s^{1/r}})\,\frac{s^n}{[n]_{q^{r}}!}
 \sum_{k = 0}^{\infty}(q^{r})^{\binom{k}{2}}\, \left[\begin{matrix}{n} \\
{k}\end{matrix}\right]_{q^{r}}
 \left(\frac{|t|}{s}\right)^{n - k} \\
 & =  \frac{s^n}{[n]_{q^{r}}!}\,\big(1 + \frac{|t|}{s}\big)^n_{q^{r}}{\mathcal{R}}_{\alpha,q}^{\delta}
 (\frac{|x|}{s^{1/r}})
 \leq  \frac{s^n}{[n]_{q^{r}}!}\,\big(1 + \frac{|t|}{s}\big)^n \, {\mathcal{R}}_{\alpha,q}^{\delta}
 (\frac{|x|}{s^{1/r}}),
\end{align*}
\text{since}, \quad $\displaystyle \frac{(q^{r})^{\delta-1
\binom{k}{2}}\,\frac{|x|^{rk}}{s^{k}}}{((r)_{q})^{rk}\prod(\alpha_{i}
+1 )_{n}^{q^{r}}}< {\mathcal{R}}_{\alpha,q}^{\delta}
(\frac{|x|}{s^{1/r}}).$
\end{proof}
\begin{lemma} \label{Lem1.4}
\[
p_{n}^{\alpha}(x, t, q^{r},\delta) \geq
\frac{\alpha_{rn,\alpha,q}}{[n]_{q^{r}}!}t^n,\quad\mbox{for } \, t,
x
> 0 ,\, \delta  > 0.
\]
\end{lemma}

\begin{proof} Since the coefficients of $p_{n}^{\alpha}$ are
positive, it follows that $$ p_{n}^{\alpha}(x, t, q^{r},\delta) \geq
p_{n}^{\alpha}(0, t, q^{r},\delta) =
\frac{\alpha_{rn,\alpha,q}}{[n]_{q^{r}}!}t^n. $$
\end{proof}

\begin{theorem} \label{thm16}
If the series $\sum_{n = 0}^{\infty}a_{n}p_{n}^{\alpha}(x_{0},
t_{0}, q^{r},\delta)$ converges for $t_{0} > 0$ and $x_{0}
> 0$, then the series $\sum_{n = 0}^{\infty}a_{n}p_{n}^{\alpha}(x_{0}, t_{0} ,
q^{r},\delta)$ and $\sum_{n = 0}^{\infty}d_{rn,\alpha,q} a_{n}p_{n -
1 }^{\alpha}(x, t , q^{r},\delta)$ converge absolutely and locally
uniformly in the strip $|t| < t_{0}$ and $\sum_{n =
0}^{\infty}a_{n}p_{n}^{\alpha}(x_{0}, t_{0} , q^{r},\delta)$ is in
${\mathcal{H}}_{\alpha}(R_{+})$ for $|t| < t_{0}$.
\end{theorem}

\begin{proof}We note by $d_{rn,\alpha,q} =
{\alpha_{rn,\alpha,q}}/{{\alpha_{r(n-1),\alpha,q}}} $. Since the
general term of a convergent series must go to zero, $\displaystyle
\lim_{n\longrightarrow \infty}a_{n}p_{n}^{\alpha}(x, t,
q^{r},\delta)= 0. $ By lemma \ref{Lem1.4}, it therefore follows that
$\displaystyle a_{n} =
O\big(\frac{[n]_{q^{r}}!}{\alpha_{rn,\alpha,q}t_{0}^n}\big)\,. $
Using Lemma \ref{Lem1.3}, we get for $s > 0$ and $\delta  \geq 1 $

\begin{align*}
&\sum_{n = 0}^{\infty}a_{n}d_{rn,\alpha,q}p_{n - 1}^{\alpha}(x, t,
q^{r},\delta)  \leq M \sum_{n = 1}^{\infty}\frac{[n]_{q^{r}}!}
{\alpha_{rn,\alpha,q}t_{0}^n}\,\frac{\alpha_{rn,\alpha,q}}{[n]_{q^{r}}!}\,(s
+ |t|)^n{\mathcal{R}}_{\alpha,q}^{\delta} (\frac{|x|}{s^{1/r}})
\\
 & \leq  M{\mathcal{R}}_{\alpha,q}^{\delta}(\frac{|x|}{s^{1/r}})\sum_{n = 0}^{\infty}\big(
 \frac{s + |t|}{t_{0}}\big)^n\, ,
\end{align*}
which converges for $s + |t| < t_{0}$. Since $s > 0$ is arbitrary it
converges for $(s + |t|) < t_{0}$, and as before for $|t| < t_{0}$.
\end{proof}

Let $f(x) =\sum_{n = 0}^{\infty}a_{n}x^n $ be an entire function of
order $\rho$, $\rho > 0$, and of type $0 < \sigma < \infty$. The
type is determined by $\displaystyle \limsup_{n \to \infty}
\frac{rn}{e\rho}\,|a_{n}|^{\frac{\rho}{rn}} = \sigma  .$ Therefore,

\begin{equation}
|a_{n}| \leq M\left(\frac{e\sigma\rho}{rn}\right)^{rn/\rho}.
\label{Eq10.9}
\end{equation}

\begin{theorem}
If $f(z) $ is an entire function of order $\rho$ with $0 < \rho <
r/r-1$ and of type $\sigma$, $0 < \sigma < \infty$, then
\begin{equation}
u(x, t) = \sum_{n = 0}^{\infty} a_{n}p_{n}^{\alpha}(x, t,
q^{r},\delta) \label{Eq10.10}
\end{equation}
is in ${\mathcal{H}}_{\alpha}(R)$ in the strip $|t| < 1/(\sigma
\rho)^{r/\rho}$ and $u(x, 0) = f(x)$.
\end{theorem}
\begin{proof}
Using \eqref{Eq10.9} and lemma \ref{Lem1.3}, for $s> 0$ we obtain

\begin{equation}
\sum_{n = 0}^{\infty} a_{n}p_{n}^{\alpha}(x, t, q^{r},\delta) \leq M
\sum_{n = 0}^{\infty} \left(\frac{e\sigma\rho}{rn}\right)^{rn/\rho}
\frac{\alpha_{rn,\alpha,q}}{[n]_{q^{r}}!}\, (s +
|t|)^n{{\mathcal{R}^{\delta}}_{\alpha,q}}(\frac{|x|}{s^{1/r}}).
\label{Eq10.11}
\end{equation}

Since, $\displaystyle
\left(\frac{e\sigma\rho}{rn}\right)^{rn/\rho}\frac{\alpha_{rn,\alpha,q}}{[n]_{q^{r}}!}\,
\leq
\left(\frac{e\sigma\rho}{rn}\right)^{rn/\rho}r^{rn}\prod_{i=1}^{r-1}
\frac{\Gamma_{q^{r}} (\alpha_{i} + n+1)}{\Gamma_{q^{r}}(\alpha_{i}
+1)}, $

or for $n\uparrow \infty$, we have $\displaystyle \prod_{i=1}^{r-1}
\frac{\Gamma_{q^{r}} (\alpha_{i} + n+1)}{\Gamma_{q^{r}}(\alpha_{i}
+1)}\sim \prod_{i=1}^{r-1} {\Gamma_{q^{r}} (\alpha_{i} + n+1)}, $ by
\cite[p. 53]{koornwinder_90} , for $n\uparrow\infty$ $\displaystyle
 \prod_{i=1}^{r-1}
{\Gamma_{q^{r}} (\alpha_{i} + n+1)}  \leq \prod_{i=1}^{r-1} {\Gamma
(\alpha_{i} + n+1)}$. Using Stirling's formula, we get
\[
\left(\frac{e\sigma\rho}{rn}\right)^{rn/\rho}r^{rn}
\prod_{i=1}^{r-1} \Gamma (\alpha_{i} + n+1) \sim \Big[\frac{e^{1 -
\frac{r-1}{r}\rho}\, r^{\rho -1}}{n^{1 - \frac{r-1}{r}\rho + (\sum
\alpha_{i}+ \frac{r-1}{2})\rho /rn}}
\Big]^{rn/\rho}\,{(2\pi)^{\frac{r-1}{2}}(\sigma\rho)^{rn/\rho}}
\]

for $0 < \rho < \frac{r}{r-1}$. Thus the series in \eqref{Eq10.11}
is dominated by
\[
M_{t,q}{\mathcal{R}}_{\alpha,q}^{\delta}(\frac{|x|}{s^{1/r}})\sum_{n
= 0}^{\infty} \{(\sigma\rho)^{r/\rho}(s+ |t|)\}^n \, ,
\]
which converges for $(\sigma\rho)^{3/\rho}(s + |t|) < 1$. Since $s
> 0$ is arbitrary, we get absolute and local uniform convergence
for $|t| < \frac{1}{(\sigma\rho)^{3/\rho}}$. Since the order and
type of  entire function is not changed by taking derivatives, a
similar type argument shows that the derived series $ \sum_{n =
1}^{\infty}a_{n}d_{rn,\alpha,q}p_{n - 1}^{\alpha}(x,
t,q^{r},\delta), $ also converges absolutely and locally uniformly
for $|t| < \frac{1}{(\sigma\rho)^{r/\rho}}$. It follows that $u(x,
t)$ given by \eqref{Eq10.10} is in ${\mathcal{H}}_{\alpha}$ in the
stated strip.
\end{proof}

\end{document}